\documentclass[twocolumn,showpacs,preprintnumbers,amsmath,amssymb]{revtex4}
\usepackage{graphicx}
\usepackage{dcolumn}
\usepackage{bm}

\begin{document}
\widetext

\title{Upper and lower critical fields in NdFeAs(O,F) single crystals : a study by Hall probe magnetization and specific heat}

\author{Z.Pribulova$^{1,2}$, T.Klein$^{1,3}$, J.Kacmarcik$^{2,4}$, C.Marcenat$^4$, M.Konczykowski$^5$, S. L. BudÕko$^6$M. Tillman$^6$ and P. C. Canfield$^6$}
\address{$^1$ Institut N\'eel, CNRS, B.P.166, 38042 Grenoble Cedex 9, France}
\address{$^2$  Centre of  Low Temperature  Physics IEP  SAS \& FS
UPJ\v S, Watsonova 47, 043 53 Ko\v{s}ice, Slovakia   }
\address{$^3$ Institut Universitaire de France and Universit\'e Joseph
Fourier, B.P.53, 38041 Grenoble Cedex 9, France}
\address{$^4$ CEA, Institut Nanosciences et Cryog\'enie, SPSMS-LATEQS
- 17 rue des Martyrs, 38054 Grenoble Cedex 9, France}
\address{$^5$ Laboratoire des Solides Irradi\'es, Ecole Polytechnique, 
91128 Palaiseau, France}
\address{$^6$ Ames Laboratory and Department of Physics \& Astronomy, 
Iowa State University, Ames, IA 50011}
\date{\today}

\begin{abstract}
The upper and lower critical fields have been deduced from specific heat and Hall probe magnetization measurements in non-optimally doped NdFeAs(O,F) single crystals ($T_c \sim 32-35$K). The anisoptropy of the penetration depth ($\Gamma_\lambda$) is temperature independent and on the order of $4.0 \pm 1.5$. Similarly specific heat data lead an anisotropy of the coherence lenght $\Gamma_\xi \sim 5.5 \pm 1.5$ close to $T_c$. Our results suggest the presence of rather large thermal fluctuations and to the existence of a vortex liquid phase over a broad temperature range ($\sim 5$K large at $2$T).
\end{abstract}

\maketitle

The recent discovery of superconductivity at unusually high temperature (up to $56$K) in rare earth iron oxypnictides \cite{Kamihara,Wang} has been the focus of a tremendous number of  theoretical and experimental works  in the past few months. The possible coexistence of superconductivity with  a complex magnetic structure \cite{Cruz} makes this system particularly fascinating and a non conventionnal pairing mechanism (associated with multi-gap superconductivity in electron and hole pockets) has been suggested by different groups \cite{Boeri}. In this context, it is very important to have a precise determination of the temperature dependence of the critical fields and corresponding anisotropies. 

We will focus on the superconducting properties of the Nd(O,F)FeAs compound  \cite{Ren1}. The aim of this paper is to present  combined Hall probe magnetization and specific heat ($C_p$) measurements performed on Nd doped single crystals. A superconducting anomaly was clearly visible in $C_p$ around $35$ K and, as previously observed in high $T_c$ oxides, this anomaly is symmetric, broad and rapidly collapses as the magnetic field is increased. On the other hand, the lower critical field presents a classical temperature dependence for both $H_a \|c$ and $H_a \|ab$ (with $H_{c1}^c(0) \sim 120 \pm 30$G and $H_{c1}^{ab} \sim  40\pm 10$ G) and the anisotropy of the penetration depth remains on the order of $4$ on the entire temperature range. A similar anisotropy is obtained for the upper critical field from the $C_p$ measurements.

\begin{figure}
\vskip 0.5cm
\includegraphics [width=7cm]{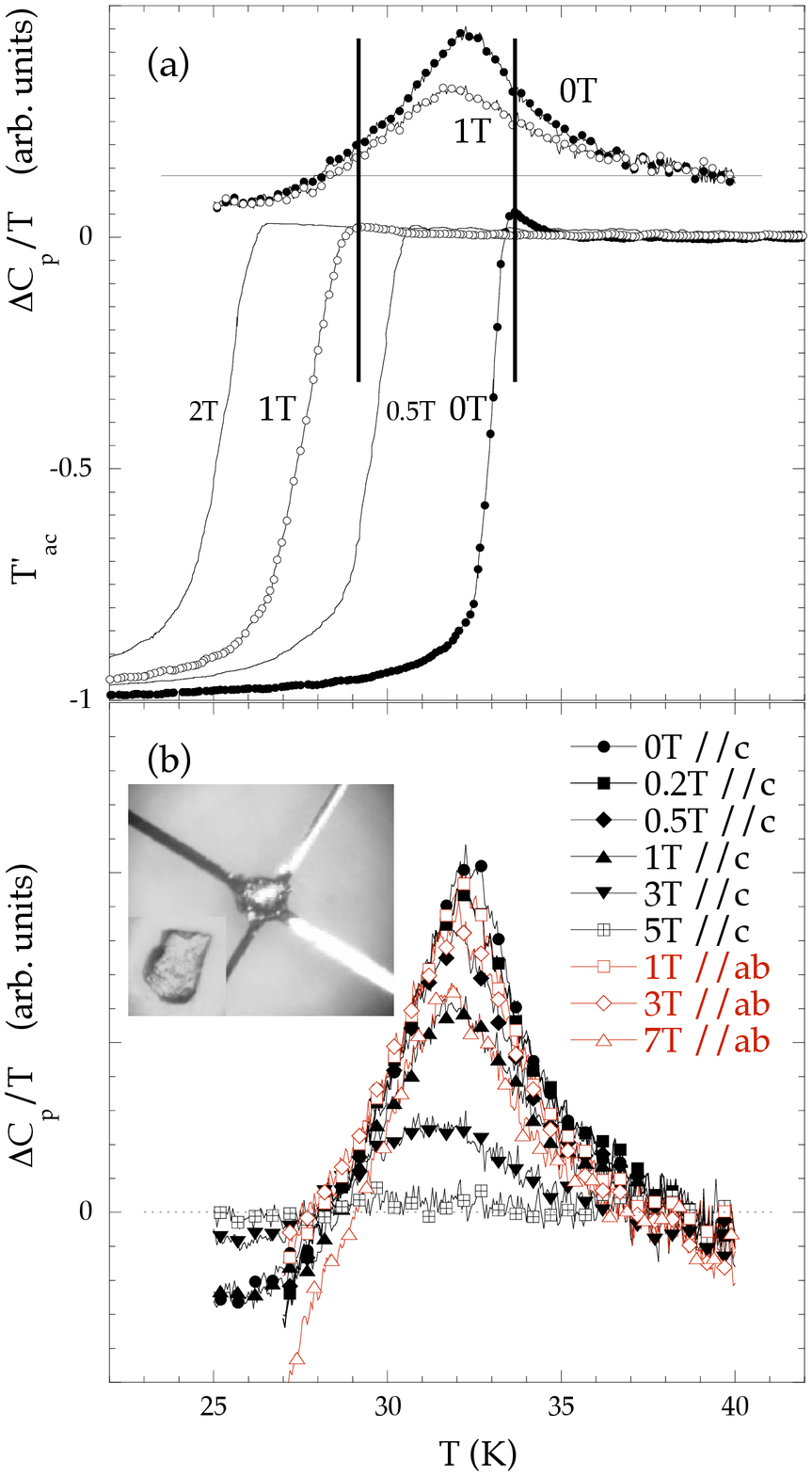}
\vskip 0.5cm
\caption{(color online) (a) Temperature dependence of the AC transmittivity and specific heat for the indicated field values ($H_a \|c$). (b) Temperature dependence of the specific heat  in a Nd(F,O)FeAs platelet (sample 1) for $H_a \|c$ (closed black symbols) and $H_a \| ab$ (open red symbols) for the indicated field values. The inset shows sample 1 mounted on the thermocouple cross, the width of the legs is $\approx 50\mu$m as well as sample 3 (bottom left corner). }
\end{figure}

Nd(O,F)FeAs samples have been synthetized at high pressure in a cubic,
multianvil apparatus. More details of the synthesis are given elsewhere \cite{Tillman}. Both Hall probe magnetization and $C_p$ measurements have been performed on the same platelet-like single  crystals extracted from the polycrystalline batch [with dimensions : $\sim100\times100\times30$  $\mu m^3$ (sample 1), $\sim120\times80\times30$ (sample 2) and $\sim170\times100\times50$ (sample 3)].

The local magnetization of the platelet has been measured by placing the sample on the top of an array of miniature Hall probes of dimensions $4 \times 4$ or $8 \times 8 \mu$m$^2$. The AC transmittivity ($T'_{ac}$, related to the {\it local susceptibility}) has been measured by applying a small ($\sim 1$G) AC field and recording the corresponding response of the probe as a function of the temperature (see Fig.1a). $T'_{ac}$ is obtained by subtracting the response in the normal state from the data and is rescaled to $-1$ at $T=4.2$K (due to the small but non zero distance between the probe and the sample surface about $10\%$ of the external field is still picked up by the probe in the superconducting state).  As shown in Fig.1a, the transmittivity measured in the center of the sample first presents a small paramagnetic increase starting at $\sim 37$K followed by a sharp diamagnetic jump at $T \sim 34$K (sample 1). This paramagnetic bump (also observed in zero field cooled DC measurements) reflects a non homogeneous distribution of the field between $34$ and $37$ K \cite{Avraham}. However the presence of a  clear anomaly in the specific heat  emphasizes the overall good quality of the platelets. A similar bump has been observed in sample 2 between $34.5$K and $37.5$K and in sample 3 between $36.5$K and $37.5$ K. Note also that previous global magnetization measurements \cite{Prozorov} yield to an onset of the diamagnetic response  of the polycrystalline batch around $T \sim 51$K in agreement with a resistivity $T_c$ on the order of $50$K for optimally doped samples \cite{Ren1}. We did not find any sign of superconductivity around $51$K in our platelets indicating that the diamagnetic response observed above $40$K in global measurements is due to a distribution of $T_c$ values (i.e. F content) within the polycristalline batch.

Specific heat measurements have been performed on the same crystallites (sample 1 and 2) using an AC technique. This high sensitivity technique  (typically $1$ part in $10^4$)  is very well adapted to measure $C_p$ of very small samples. Heat
was supplied to the sample at a frequency $\omega = 10$ Hz by a light emitting diode via an optical fiber. The temperature oscillation have been recorded with a thermocouple which has been calibrated from measurements on ultra pure silicon. The superconducting contribution to the specific heat ($\Delta C_p$) has been obtained by subtracting the curve at $\mu_0H_a=7T$ (for $H_a \| c$) from the curves obtained for lower fields as well as a $(H_a/T)^2$ contribution to account for the presence of a magnetic background . As shown in Fig.1b (sample 1),  there is no sharp discontinuity at $T_c$ and the anomaly has the shape of a rounded peak with a height on the order of a few $10^{-3}$ of the total specific heat. The other striking behaviour is the influence of the magnetic field. As previously obsersed in cuprates \cite{Marcenat}, the smearing of the anomaly with field is unexpectedly strong with an almost field independent onset. The maximum of $C_p$ is pushed downwards in temperature by the field with a characteristic average slope on the order of $-2$ T/K for $H_a\|c$ and $-7$ T/K for $H_a\|ab$) [compared to resp. $-1.5$ T/K and $-9$ T/K in optimally doped YBaCuO, note that the shift is possibly non linear for small magnetic field, see solid line in Fig.2 and discussion below]. As in cuprates, the shape of the anomaly precludes any precise determination of $H_{c2}$. However,  assuming that $H_{c2}$ is close to the maximun of $C_p$ ($H_{max}$) and taking the corresponding  linear temperature dependence, we hence get $H_{c2}(0) \sim 0.7 \times dH_{max}/dT \times T_c \sim 40$T and $\sim  170$ T for $H_a\|c$ and $H_a\|ab$ respectivevely. The corresponding anisotropy of the coherence length $\Gamma_\xi = \xi_c/\xi_{ab} = \Gamma_{H_{c2}}$ is hence on the order of  $H_{c2}^{ab}(0)/H_{c2}^c(0) \sim 4$. The uncertainty on the $H_{c2}$ values is rather large but $\Gamma_\xi$ can also be estimated from the total shape of the anomaly writting : $\Delta C(\Gamma_\xi \times H_a\|ab) \sim \Delta C(H_a\|c)$ consistently leading to $\Gamma_\xi \sim 5.5 \pm 1.5$.  This value is also in good agreement with the one previously reported by Welp {\it et al.} \cite{Welp} and with recent band structure calculations \cite{Singh}.  \

The influence of $H_a$ on the $C_p$ anomaly and its overall shape is strikingly different from the one observed is classical superconductors or even in the (K,Ba)BiO$_3$ system ($T_c \sim 32K$) \cite{KleinKBBO} or MgB$_2$ ($T_c \sim 39$ K) \cite{LyardCp} of similar $T_c$ values. It is tempting to attribute these deviations to fluctuation effects. To reinforce this idea, it is worth noting that the onset of the diamanetic response well coincides with the inflection point of the $C_p$ anomaly for $H_a = 0$ but is pushed towards substantially lower temperature in field ($\sim 5K$ lower than the $C_p$ shift for $\mu_0H_a = 2$T). As $T'_{ac}$ is sensitive to pinning, this result suggests that the irreversibility line is well separated from the superconducting transition for $H_a \neq 0$ (see inset of Fig.2 for for $H_a \| c$, a similar result - not shown - is obtained for $H_a \| ab$) supporting the idea of an extended vortex liquid phase. The importance of thermal fluctuations can be quantified by the Ginzburg number \cite{Blatter} $G_i = (k_BT_c/\epsilon_0\xi_c)^2/8$ where $\epsilon_0$ ($=(\Phi_0/4\pi\lambda_{ab})^2$) is the line tension of the vortex matter. Large $\lambda$ values (see below) combined with small $\xi$ values hence lead here to $\epsilon_0\xi_c \sim 200$ K  (i.e. similar to cuprates) compared to $\sim 1000$ K in (K,Ba)BiO$_3$ and even $\sim 10^4$ in MgB$_2$. Although smaller than in cuprates  due to smaller $T_c$ values, thermal fluctuations are hence expected to play a role in this system possibly leading to the melting of the vortex solid  ($G_i \sim 3\times10^{-3}-10^{-2}$). A similar conclusion has been drawn by Welp {\it et al.} \cite{Welp}.  Note that, as expected for vortex melting, the irreversibility line clearly presents a positive curvature ($H_{irr} \propto (1-T/T_c)^\alpha$ with $\alpha \sim 1.5$, for a review see \cite{Blatter}). Similarly, thermal fluctuations may also induce an  upward curvature in the $H_{c2}$ line \cite{Klein} but eventhough our data suggest the existence of such a curvature   (see also \cite{Welp}), we cannot unambiguously conclude given the uncertanties and limited temperature range of the $C_p$ data.

\begin{figure}
\vskip 0.5cm
\includegraphics [width=7cm]{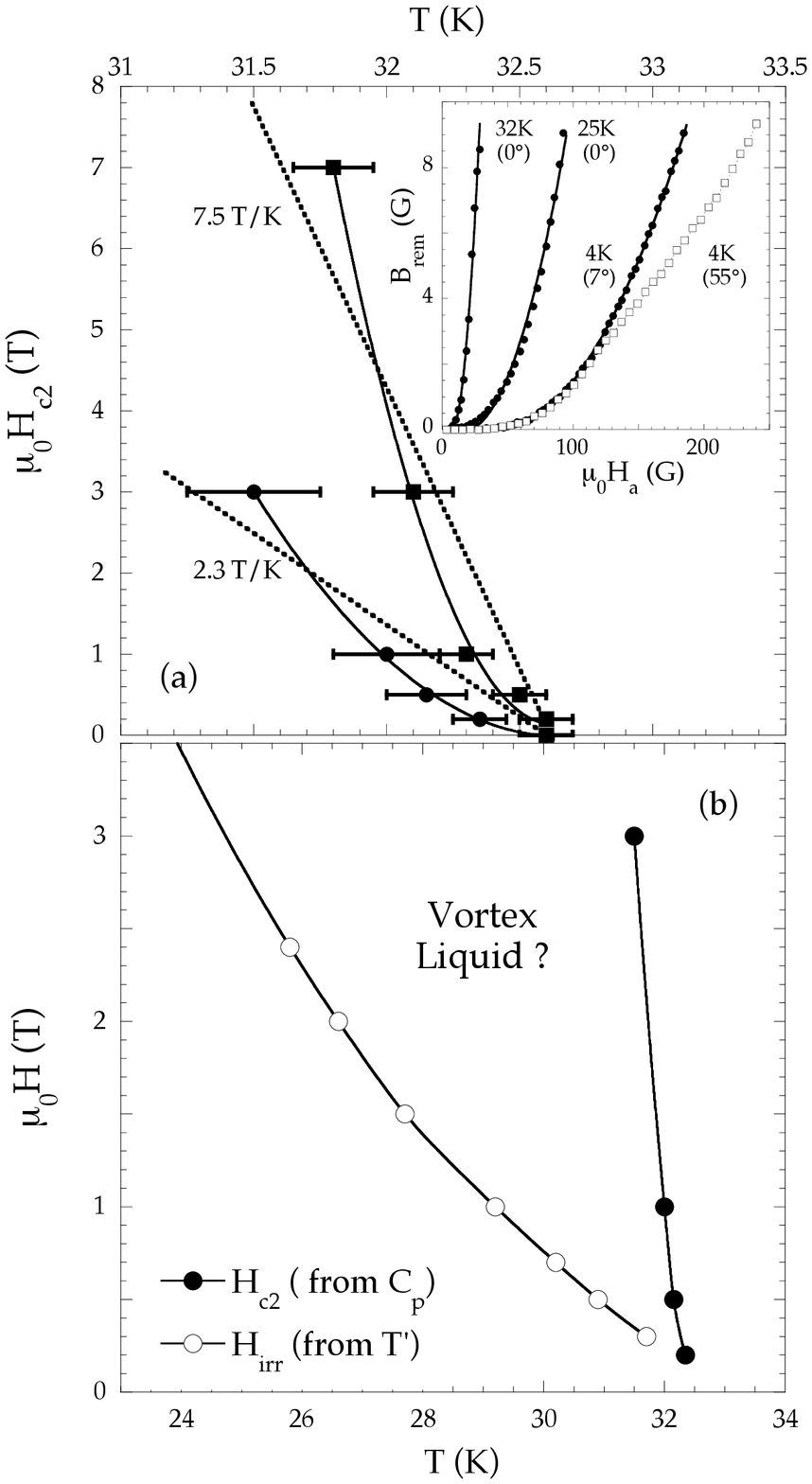}
\vskip 0.5cm
\caption{(a) Temperature dependence of the upper critical field (maximum of the specific heat anomaly, see Fig.1a) for $H_a \|c$ (circles) and $H_a \|ab$ (squares). In the inset :  Remanent field ($B_{rem}$) as a function of the applied field $\mu_0H_a$ (see text for details) for the indicated temperatures and field orientation. The solid lines are $(H_a-H_p)^2$ fits to the data. (b) : comparison between $H_{c2}$ (deduced from $C_p$, closed symbols) and the onset of the diamagnetic response $H_{irr}$ (open symbols) for $H_a \|c$ suggesting the existence of a broad vortex liquid phase.}
\end{figure}

\begin{figure}
\vskip 0.5cm
\includegraphics [width=8cm]{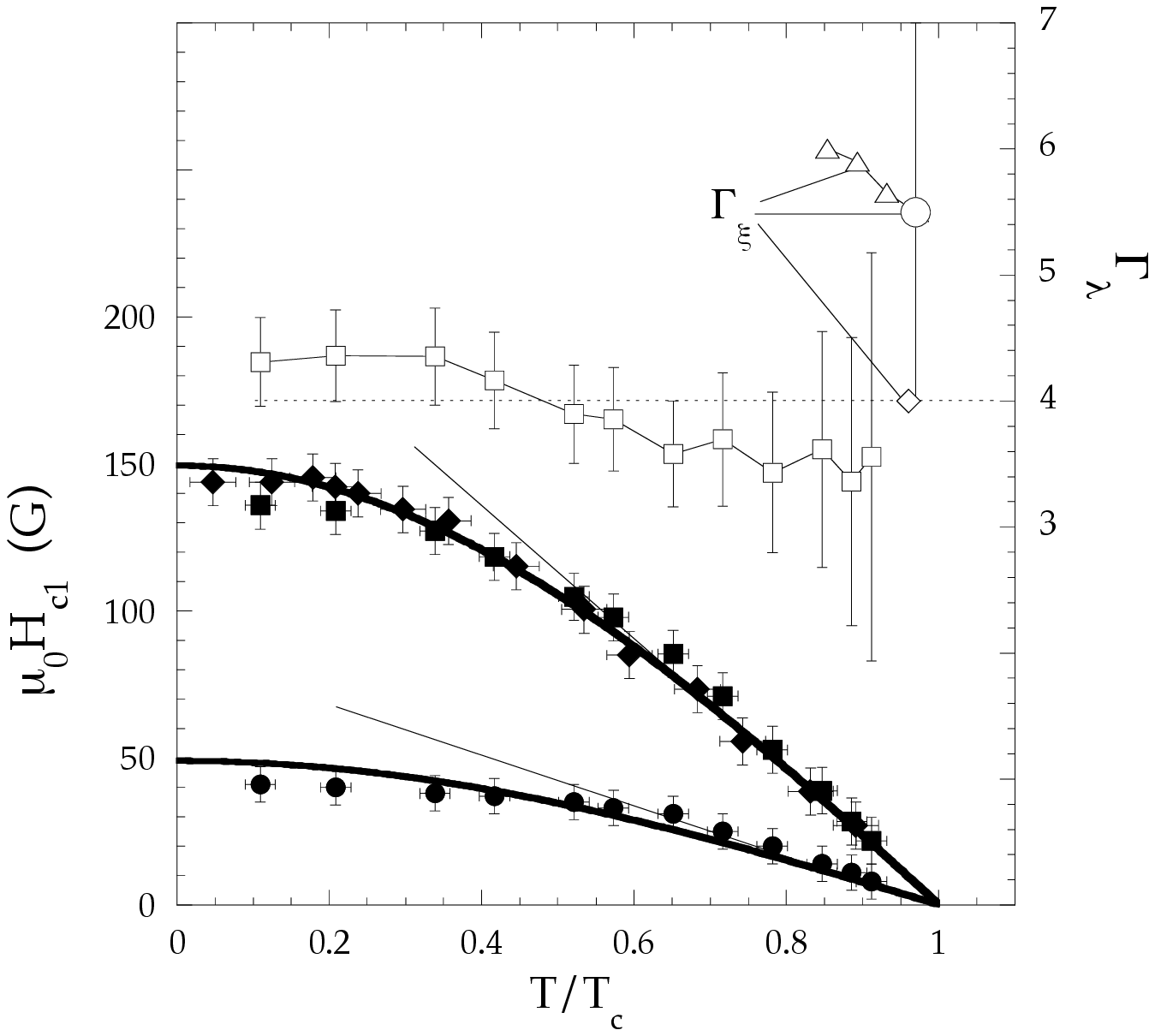}
\vskip 0.5cm
\caption{Temperature dependence of the lower critical field $H_{c1}$ along the $c$ direction (solid squares for sample 3, solid lozanges for sample 1, rescaled by a factor of 1.5) and ab plane for sample 3 (solid circles). The solid lines are the standard behaviours expected from the the BCS theory. The corresponding temperature dependence of the anisotropy ($\Gamma_\lambda =1.3\times\Gamma_{H_{c1}}$) is also reported (open squares). The error bars correspond to the uncertainties on the determination of $H_p$, the whole curve might be shifted $\pm 1.5$ due to the uncertainty of the demagnetization factor. In all cases, $\Gamma_\lambda$ remains flat in T. The $\Gamma_\xi$ values deduced from $C_p$ (present work : open circle and ref [13] : open lozange) and transport : open triangles (from ref. [26]) measurements are also displayed.}
\end{figure}

The first penetration field $H_p$ has been deduced by measuring the remanent field ($B_{rem}$) in the sample after applying an external field $H_a$ and sweeping the field back to zero. For $H_a <H_p$ no vortices penetrate  the sample and the remanent field remains equal to zero. $H_a$ is progressively increased until a finite remanent field is obtained as vortices remain pinned in the sample for $H_a > H_p$ ($B_{rem}$ increasing as $(H_a-H_p)^2$, see solid line in the inset of Fig.1a). Theoretically speaking, $H_p$ is expected to depend on the position of the probe, being larger in the center of the sample as vortices first remain pinned close to sample edges. However, experimentally, a non-zero $B_{rem}$ value could be detected almost simultaneously on all the probes due to the non zero distance between the sample and the probe.  Note that we did not observe any significant change in the $B_{rem}(H_a)$ curve up to $H_a \sim 3 H_p$ as the magnetic field was tilted away from the c-axis (with an angle $\theta_{H_a} \leq 60-70^\circ$). Even though, the angle between the internal field ($\theta_H$) and the c-axis is reduced by the demagnetization effects,  this surprizing behaviour suggests that the induction ($B$) in the sample remains perpendicular to the platelets up to large $\theta_H$ values (see inset of Fig.1a for sample 1 at $\theta_{H_a} = 7^\circ$ and $55^\circ$).  $B_{rem}$ finally decreases with $\theta$ for $H_a \geq 3H_p$ (and/or for  $\theta_{H_a} \geq 70^\circ$), indicating that the vortices finally aligns on the applied field for large $H_a$ (and/or $\theta_{H_a}$) values. 

In samples with rectangular cross sections,  flux lines partially penetrate into the sample through the sharp corners even for $H_a < H_p$ \cite{Brandt,Zeldov} but remain "pinned" at the sample equator. The magnetization at $H_a=H_p$ is then larger than $H_{c1}$ and the standard "elliptical" correction for $H_{c1}$ ($=H_p/(1-N)$ where $N$ is the demagnetization factor) can not be used anymore.  Following \cite{Brandt}, in presence of geometrical barriers, $H_{p}$ is related to $H_{c1}$ through $H_{c1} \approx H_p/tanh(\sqrt{\alpha d / 2w})$ where $\alpha$ varies from 0.36 in strips to 0.67 in disks ($2w$ and $d$ being the sample width and thickness, respectively).Taking an average $\alpha$ value $\sim 0.5$ \cite{Geombarriers} we hence got \cite{elliptical} : $H_{c1}^c(0) \sim 110 \pm 20$G in sample 1, $\sim 100 \pm 20$G in sample 2 and $H_{c1}^c(0) \sim 140 \pm 30$G in sample 3.  To obtain $H_p$ in the $ab$- plane, sample 3 has been rotated by $90^\circ$ putting the probe on the small side of sample (i.e. applying $H_a$ along the long direction of the sample so that $H_p^{ab}(0) \sim H_{c1}^{ab}$). We hence obtained $H_{c1}^{ab} \sim 40 \pm 10$G (in sample 3). 

The temperature dependence of $H_{c1}$ for both $H_a \| c$ (sample sample 1 and 3; the data for sample 1 have been rescaled by a factor $1.5$) and $H_a \| ab$ (sample 3) is displayed in Fig.3. As shown, $H_{c1}^c$ clearly flattens off at low temperatures (down to $1.6$K for sample 1) indicating that the gap is fully open in our Nd(O,Fe)AsFe single crystals. This result is in striking contrast with $H_p$ measurements by  \cite{Ren2} which suggested that $H_{c1}$ varies linearly at low temperature in La(O,F)FeAs hence suggesting the existence of gapless superconductivity. On the other hand, a very similar temperature dependence (and $H_{c1}(0)$ values) has been recently obtained by Okazaki {\it et al.} in Pr doped samples \cite{Okazaki}.

The lower critical field is related to the penetration depth through : $H_{c1}^c=\Phi_0^2/(4\pi\lambda_{ab}^2)(Ln(\kappa)+c(\kappa))$ where $\kappa=\lambda_{ab}/\xi_{ab}$ ($\xi$ being the coherence length) and $c(\kappa)$ a $\kappa$ dependent function tending towards $\sim 0.5$ for large $\kappa$ values. Taking $H_{c2}^c(0) \sim 40$T, one gets $\lambda_{ab}(0) \sim 270 \pm 40$ nm \cite{erreurHc2}. Those $\lambda$ values are consistent with those previous obtained in oxypnictides. Indeed, slightly lower $\lambda_{ab}$ values ($\sim 200$ nm) have been reported in both Sm and Nd doped samples with higher $T_cs$ ($ \sim 50$K, from either torque \cite{Weyeneth}, muons relaxation  \cite{Drew} or plasma frequency \cite{Dubroka} measurements) whereas $\lambda \sim 360-390$nm have been obtained in La doped samples with lower $T_c$ values ($\sim 23$K, \cite{Luetkens,Ren2}).

For $H \| ab$,  $H_{c1}^{ab}=\Phi_0^2/(4\pi\lambda_{ab}\lambda_c)(Ln(\kappa^*)+c(\kappa^*))$ with $\kappa* = \lambda_c/\xi_{ab}$, and $H_{c2}^{ab}=\Phi_0^2/(2\pi\xi_c\xi_{ab}) \sim 170$ T, leading to $\lambda_c \sim 1200 \pm 200$ nm and a corresponding anisotropy ($\Gamma_\lambda$) on the  order of $4.0$ ($\pm 1.5$), almost temperature independent down to $1.6$K (see Fig.3, sample 3, $\Gamma_\lambda = \Gamma_{H_{c1}}\times(1+Ln(\Gamma_\lambda)/(Ln(\kappa)+0.5)) \sim 1.3\times\Gamma_{H_{c1}}$). Note that our data are in striking contrast with those deduced from torque measurements  in Sm doped samples \cite{Weyeneth} which led to strongly increasing $\Gamma_\lambda$ values, exceeding $30$ at low temperature. On the other hand, our $\Gamma_\lambda$ consistently tends towards $\Gamma_\xi$ for $T \rightarrow T_c$ deduced either from $C_p$ (present work and \cite{Welp}) or transport data \cite{Jia}. The difference between the anisotropies measured at low ($\Gamma_\lambda$) and high ($\Gamma_{\xi}$) fields remain within our error bars and we hence cannot comment on a possible similarity with MgB$_2$ in which the coexistence of 2 gaps leads to a strongly field and temperature dependence of the anisotropy \cite{Lyard}. 

In conclusion, specific heat and Hall probe magnetization data suggest than $\Gamma_\xi \sim \Gamma_\lambda \sim 5 \pm 2$ in Nd(F,O)FeAs single crystals. Whereas the irreversibility is rapidly shifting towards lower temperature for increasing $H_a$, the $C_p$ anomaly collapses and shows only a minor shift suggesting the presence of strong thermal fluctuations possibly leading to the melting of the vortex solid.

We would like to thank K.van der Beek for very fruitful discussions and preliminary magneto-optic images. Work at the Ames Laboratory was supported by the Department of Energy, Basic Energy Sciences under Contract No. DE-AC02-07CH11358. Z.P. thanks to the Slovak Research and Development Agency under the contract No. LPP-0101-06, J.K. thanks to 6th Framework Programme MCA Transfer of Knowledge project ExtreM No. MTKD-CT-2005-03002.


\begin{references}
\bibitem{elliptical}  Note that for the rather high $d/2w$ ratio of our platelets  a standard "elliptical" correction would only slightly overestimate $H_{c1}$ : for sample 1 and 2 $d/2w\sim 0.3$ leading to $1/(1-N) \sim 4.0 \pm 0.2$ for strips and $2.8 \pm 0.2$ for disks compared to $H_{c1}/H_p \sim 2.7 \pm 0.4$ for geometrical corrections.
\end{references}
\end{document}